\documentclass[fleqn,twoside]{article}
\usepackage{espcrc2}

\usepackage{graphicx}

\title{Gamma-ray bursts: Potential sources of ultra high energy cosmic-rays}

\author{E. Waxman\address[WIS]{Physics Faculty, Weizmann Institute of
        Science, Rehovot 76100, Israel\\ waxman@wicc.wiezmann.ac.il}
        \thanks{To appear in Nuclear Physics B (Proceedings Supplement),
        Proc. XIII International Symposium on 
        Very High Energy Cosmic Ray Interactions (Pylos Greece, September 2004)}}

\begin{document}

\begin{abstract}

The arguments suggesting an association between the sources of cosmological gamma-ray bursts (GRBs) and the sources of ultra-high energy cosmic rays (UHECRs) are presented. Recent GRB and UHECR observations are shown to strengthen these arguments. Predictions of the GRB model for UHECR production, that may be tested with large area high energy cosmic-ray detectors which are either operating or under construction, are outlined.

\end{abstract}
\maketitle

\section{Introduction}
\label{sec:introduction}

The origin of UHECR's is one of the most exciting open questions of high energy astrophysics \cite{Sigl,NaganoWatson}. The extreme energy of the highest energy events poses a challenge to models of
particle acceleration. Since very few known astrophysical objects have characteristics indicating that they may allow acceleration of particles to the observed high energies \cite{pascos03}, the question of whether GRBs are possible UHECR sources is of great interest. Moreover, since the GRB model for UHECR production makes unique predictions, which differ from those of other models \cite{W01rev}, the design and analysis of future UHECR experiments, as well as of high energy neutrino and photon experiments, may be affected by the answer to this question.

The phenomenology of GRBs, bursts of 0.1 MeV--1 MeV photons lasting for a few seconds \cite{Fishman95}, suggests that the observable effects are due to the dissipation of the kinetic energy of a cosmologically distant, relativistically expanding wind, a ``fireball,'' whose primal cause is not yet known \cite{fireballs1,fireballs2,fireballs3}. Waxman \cite{W95a}, Milgrom \& Usov \cite{MnU95} and Vietri \cite{Vietri95} have independently suggested that ultra-high energy, $>10^{19}$~eV, cosmic rays may be produced in GRB sources. The model suggested in ref. \cite{W95a} was based on two arguments. First, it was shown that the constraints imposed on the relativistic wind by the requirement that it produces observed GRB
characteristics are similar to the constraints imposed on such a wind by the requirement that it would allow proton acceleration to $>10^{20}$~eV. Second, the energy generation rate of $\gamma$-rays by GRBs was shown to be similar to the energy generation rate required to account for the observed UHECR flux \cite{W95a,W95b}. 

In \S~\ref{sec:model} we briefly describe the model proposed in ref.~\cite{W95a} for UHECR production in GRB fireballs. In \S~\ref{sec:recent} we discuss the implications of recent GRB and UHECR observations to this model (for a more detailed discussion, see \cite{W04}). In \S~\ref{sec:predictions} we discuss predictions of the model that may be tested with high energy cosmic-ray experiments. For predictions of the model that may be tested with high energy neutrino detectors see \cite{W01rev,PM04rev,HalzenRev} and references therein. Our conclusions are discussed in \S~\ref{sec:conclusions}.

\section{Production of UHECRs in GRBs}
\label{sec:model}

\subsection{Proton acceleration}
\label{sec:acceleration}

General phenomenological considerations, based on $\gamma$-ray
observations, indicate that, regardless of the nature of the
underlying sources, GRB's are produced by the dissipation of the
kinetic energy of a relativistic expanding fireball. A compact
source, $r_0\sim10^7$~cm, produces a wind, characterized by an
average luminosity $L\sim10^{52}{\rm erg\ s}^{-1}$ and mass loss
rate $\dot M$. At small radius, the wind bulk Lorentz factor,
$\Gamma$, grows linearly with radius, until most of the wind
energy is converted to kinetic energy and $\Gamma$ saturates at
$\Gamma\sim L/\dot M c^2\sim300$. Variability of the source on a
time scale $\Delta t\sim10$~ms, resulting in fluctuations in the
wind bulk Lorentz factor $\Gamma$ on a similar time scale, results
in internal shocks in the ejecta at a radius $r\sim
r_d\approx\Gamma^2c\Delta t\gg r_0$. It is assumed that internal
shocks reconvert a substantial part of the kinetic energy to
internal energy, which is then radiated as $\gamma$-rays by
synchrotron and inverse-Compton radiation of shock-accelerated
electrons. At a later stage, the shock wave driven into the
surrounding medium by the expanding fireball ejecta leads to the
emission of the lower-energy afterglow.

The observed radiation is produced, both during the GRB and the
afterglow, by synchrotron emission of shock accelerated electrons.
In the region where electrons are accelerated, protons are also
expected to be shock accelerated. Thus, it is likely that protons, as well as
electrons, are accelerated to high energy within GRB fireballs.

The internal shocks within the expanding wind are expected to be mildly
relativistic in the wind rest frame, due to the fact that the allowed range of Lorentz
factor fluctuations within the wind is from few~$\times10^2$ (the lower limit required to avoid large optical depth) to few~$\times10^3$ (the maximum Lorentz factor to which shell acceleration by radiation pressure is possible, e.g. \cite{fireballs3}). This implies that the Lorentz factors associated
with the relative velocities are not very large. Since internal shocks are mildly
relativistic, we expect our understanding of non-relativistic shock acceleration to apply to
the acceleration of protons in these shocks. In particular, the predicted energy distribution of accelerated protons is expected to be $dn_p/d\epsilon_p\propto \epsilon_p^{-2}$, similar to the electron energy spectrum inferred from the observed photon spectrum. 

A power law energy spectrum with index 2 has
been observed for non-relativistic shocks (see, e.g., ref. \cite{non-relativistic} and references therein) and for relativistic shocks~\cite{relativistic}. It is consistent with that expected theoretically for Fermi acceleration in collisionless shocks~\cite{non-relativistic,rel-theory}, although a first principles understanding of the process is not yet available (see, e.g. ref.~\cite{arons} for a discussion of alternative shock acceleration processes).

Several constraints must be satisfied by wind parameters in order
to allow proton acceleration to high energy $\epsilon_p$. We summarize
below these constraints. The reader is referred to refs.~\cite{W95a,W01rev} for a detailed derivation. The requirement that
the acceleration time, assumed to be comparable to the Larmor gyration time of the accelerated particle, be smaller than the wind expansion time, which also implies that the proton is confined to the
acceleration region over the required time, sets a lower limit to
the strength of the wind magnetic field. This may be expressed as
a lower limit to the ratio of magnetic field to electron energy
density \cite{W95a},
\begin{equation}
u_B/u_e>0.02 \Gamma_{2.5}^2 \epsilon_{p,20}^2L_{\gamma,52}^{-1},
\label{eq:xiB}
\end{equation}
where $\epsilon_p=10^{20}\epsilon_{p,20}$~eV, $\Gamma=10^{2.5}\Gamma_{2.5}$ and
$L_{\gamma}=10^{52}L_{\gamma,52}{\rm erg/s}$ is the wind
$\gamma$-ray luminosity. A second constraint is imposed by the
requirement that the proton acceleration time be smaller than the
proton energy loss time, which is  dominated by synchrotron
emission. This sets an upper limit to the magnetic field strength,
which in turn sets a lower limit to $\Gamma$ \cite{W95a,RnM98}
\begin{equation}
\Gamma>130 \epsilon_{p,20}^{3/4}\Delta t^{-1/4}_{-2}. \label{eq:Gmin}
\end{equation}
Here, $\Delta t=10^{-2}\Delta t_{-2}$~s. As explained in
\cite{W95a}, the constraints Eq.~(\ref{eq:xiB}) and Eq.~(\ref{eq:Gmin})
hold regardless of whether the fireball is a sphere or a narrow
jet (as long as the jet opening angle is $>1/\Gamma$). The
luminosity in Eq.~\ref{eq:xiB} is the "isotropic equivalent
luminosity", i.e. the luminosity under the assumption of isotropic
emission.

Internal shocks within the wind take place at a radius $r_d\approx\Gamma^2c\Delta t$. The constraint of Eq.~(\ref{eq:xiB}) is independent of $\Delta t$, i.e. independent of the internal collision radius, while the constraint of Eq.~(\ref{eq:Gmin}) sets a lower limit to the collision radius for a given $\Delta t$. This implies that protons may be accelerated to $>10^{20}$~eV regardless of the value of $\Delta t$, which may range from the dynamical time of the source ($\Delta t\sim1$~ms) to the wind duration ($\Delta t\sim1$~s), provided the magnetization and Lorentz factor are sufficiently large, following Eqs.~\ref{eq:xiB} and~\ref{eq:Gmin}. 

At large radii the external medium affects fireball evolution, and a "reverse shock" is driven backward into the fireball ejecta and decelerates it. For typical GRB fireball parameters this shock is also mildly relativistic (e.g. \cite{fireballs3}), and its parameters are similar to those of an internal shock with $\Delta t\sim10$~s. Protons may therefore be accelerated to $>10^{20}$~eV not only in the internal wind shocks, but also in the reverse shock \cite{WnB-AG,W01rev}. 
This implies that proton acceleration to $>10^{20}$~eV is possible, provided the constraints of Eqs.~\ref{eq:xiB} and~\ref{eq:Gmin} (with $\Delta t\sim10$~s) are satisfied, also in (the currently less favorable) scenario where GRB $\gamma$-rays are produced in the shock driven by the fireball into the surrounding gas, rather than by internal collisions, as suggested, e.g., in \cite{Dermer99}. 

The constraints given by Eqs.~(\ref{eq:xiB}) and (~\ref{eq:Gmin}) are
remarkably similar to those inferred from $\gamma$-ray
observations, based on independent physical arguments: $\Gamma>300$ is implied by the $\gamma$-ray spectrum by the requirement to avoid high pair-production optical depth, and magnetic field close to equipartition, $u_B/u_e\sim0.1$, is required in order to account for the observed $\gamma$-ray emission \cite{fireballs1,fireballs2,fireballs3}. This was the basis for the association of GRB's and UHECR's  suggested in \cite{W95a}. 

It should be noted here, that the only assumption related to the acceleration process made in the derivation of the constraints~(\ref{eq:xiB}) and~(\ref{eq:Gmin}) is that the acceleration time is comparable to the Larmor gyration time of the accelerated particle. These constraints are valid therefore not only for Fermi-type shock acceleration, but rather to any electromagnetic acceleration process. This point is illustrated, for example, by the results of the analysis of ref.~\cite{Pelletier}, where a non Fermi-type electromagnetic acceleration process is considered.

\subsection{gamma-ray and UHECR energy production rates}
\label{sec:rate}

The GRB model for UHECR production was suggested prior to the detection of GRB afterglows \cite{AG_ex_review}. Estimates of the rate of GRBs were based at that time on the $\gamma$-ray flux distribution, and ranged from $\sim3/{\rm Gpc^{3}yr}$ \cite{Piran92} to $\sim30/{\rm Gpc^{3}yr}$ \cite{MnP92}. The estimated average $\gamma$-ray energy release in a single GRB, based on a characteristic peak flux of $\sim10^{51}{\rm erg/s}$ \cite{Fishman95}, was $\sim10^{52}$~erg. These estimates were subject to large uncertainties, since the $\gamma$-ray luminosity function as well as the evolution of GRB rate with redshift were poorly constrained. Based on the rate and energy estimates, the rate of $\gamma$-ray energy generation by GRBs was estimated to be $\sim10^{44}{\rm erg/Mpc^3yr}$. This rate is similar to the energy generation rate in cosmic-rays of energy in the range of $10^{19}$~eV to $10^{21}$~eV, $4.5\pm1.5\times10^{44}{\rm erg/Mpc^3yr}$, inferred from Fly's Eye and AGASA measurements of the UHECR flux available at the time the GRB model for UHECR production was proposed \cite{W95b}. 

The determination of GRB redshifts, which was made possible by the detection of afterglows, allows a more reliable estimate of the GRB energy production rate, and the increased exposure of UHECR experiments, provided by AGASA and HiRes, allows a more accurate estimate of the UHECR energy production rate. The implications of these more recent observations is discussed in \S~\ref{sec:rate:Implications}.

\section{Implications of recent GRB and UHECR observations}
\label{sec:recent}

\subsection{Proton acceleration}
\label{sec:acc:Implications}

Afterglow observations \cite{AG_ex_review} lead to the confirmation of the
cosmological origin of GRBs and confirmed standard model
predictions of afterglow that results from synchrotron emission of
electrons accelerated to high energy in the highly relativistic
shock driven by the fireball into its surrounding gas \cite{fireballs1,fireballs2,fireballs3}. Afterglow observations provide therefore strong support for the  underlying fireball scenario. In addition, afterglow observations provide important information on the values of model parameters that enter the constraints given by Eqs.~(\ref{eq:xiB}) and (\ref{eq:Gmin}).

Prior to the detection of afterglows, it was commonly assumed that
the farthest observed GRB's lie at redshift $z\sim1$ \cite{MnP92,Piran92}. Based on afterglow redshift determinations, we now know that detected GRB's typically lie at farther distances (e.g. \cite{Bloom03}). This implies that the
characteristic GRB luminosity is higher by an order of magnitude compared to pre-afterglow estimates, $L_\gamma\approx10^{52}{\rm erg/s}$ instead of $L_\gamma\approx10^{51}{\rm erg/s}$. This relaxes the constraint on magnetic field energy fraction given by Eq.~(\ref{eq:xiB}). 

In several cases, fast follow up afterglow observations
allowed the detection of radio and optical emission from the reverse shock
(e.g. \cite{ZKM03,SR03}). These observations provide direct information on the plasma conditions in the reverse shock, where acceleration of protons
to high energy may take place. Two major conclusions were drawn from the analysis of the early optical and radio reverse shock emission. First, lower limits to the initial fireball Lorentz factors were inferred, in the range of $\Gamma>100$ to $\Gamma>1000$ \cite{ZKM03,SR03}. Second, the magnetic field in the reverse shock was inferred to be close to equipartition, that is $u_B/u_e$ was inferred to be of order unity \cite{Draine00,ZKM03}. Early afterglow observations provide therefore constraints on $\Gamma$ and on $u_B/u_e$ which are (i) Independent of the constraints derived from $\gamma$-ray observations; (ii) Consistent with the $\gamma$-ray constraints; and (iii) Are remarkably similar to the constraints of Eqs.~(\ref{eq:xiB}) and (\ref{eq:Gmin}), that need to be satisfied in order to allow proton acceleration to $>10^{20}$~eV.

\subsection{gamma-ray and UHECR energy production rates}
\label{sec:rate:Implications}

\subsubsection{gamma-ray production rate}

Most of the GRBs are observed from $z>1$, since they can be detected out to large redshift. This implies that the GRB rate density at $z>1$ is better constrained by the observations than the local, $z=0$, rate. The inferred local rate depends on the assumed redshift evolution. It is now commonly believed that the GRB rate evolves with redshift following the star-formation rate, based on the association of GRBs with type Ib/c supernovae. This association is based on the temporal and angular coincidence of several GRBs and type Ib/c supernovae \cite{Galama98,Stanek03,Hjorth03}, and on evidence for optical supernovae emission in several GRB afterglows \cite{Bloom03a}. Adopting the assumption, that the GRB rate follows the redshift evolution of the star formation rate, the local ($z=0$) GRB rate density is \cite{GPW03} $R_{\rm GRB}(z=0)\approx 0.5\times{\,10^{-9}\rm Mpc^{-3}~yr^{-1}}$. Given the current (systematic uncertainties in the redshift) data, this rate is accurate to within a factor of a few \cite{GPW03}.

The local energy generation rate in $\gamma$-rays by GRB's, $\dot{\varepsilon}_{\gamma}$, is given by the product of $R_{\rm GRB}(z=0)$ and the average $\gamma$-ray energy release in a single GRB, $\varepsilon_\gamma$. \cite{Bloom03} provide $\varepsilon_\gamma$ for 27 bursts with known redshifts, in a standard rest-frame bandpass, 0.02~MeV to 2~MeV. The average is $\varepsilon_\gamma=2.9\times10^{53}$~erg, with estimated uncertainty, due to the correction to a fixed rest-frame bandpass, of $\sim20\%$ for individual bursts (and much smaller for the average). In calculating $\dot{\varepsilon}_{\gamma}$ from this value of $\varepsilon_\gamma$, the following point should be taken into account. $\varepsilon_\gamma$ is the average energy for bursts with known redshift, most of which were localized by the BeppoSAX satellite. Since BeppoSAX has a higher detection flux threshold than BATSE \cite{Band03,GPW03}, it is sensitive to $\approx70\%$ of the bursts detectable by BATSE, for which the GRB rate $R_{\rm GRB}(z=0)$ was inferred. Thus, the energy generation rate by bursts detectable by BeppoSAX is 
\begin{eqnarray}
\nonumber
\dot{\varepsilon}_{\gamma[0.02\rm MeV,2MeV]}^{\rm GRB}&\approx&  0.7R_{\rm GRB}(z=0)\varepsilon_\gamma\\&=& 10^{44} {\rm erg~Mpc^{-3}~yr^{-1}}. \label{eq:GRB_g_rate}
\end{eqnarray}
The energy observed in $\gamma$-rays in the range of BATSE is dominated by photons in the energy range of 0.1~MeV to 1~MeV, and therefore reflects the energy of accelerated electrons over roughly half a decade of electron energies (recall that for synchrotron and inverse-Compton emission the photon energy is proportional to the square of the electron energy). Thus, the rate of energy generation per logarithmic interval of electron energy is 
\begin{eqnarray}
\nonumber\epsilon_e^2\frac{d\dot{n}_e^{\rm GRB}}{d\epsilon_e}&\approx&\dot{\varepsilon}_{\gamma[0.02\rm MeV,2MeV]}^{\rm GRB}\\ &\approx&10^{44}{\rm erg~Mpc^{-3}~yr^{-1}}. \label{eq:GRB_e_rate}
\end{eqnarray}

The main uncertainty in determining $\dot{\varepsilon}_{\gamma}$ is related to the uncertainty in the local GRB rate, due to which the value given in Eq.~\ref{eq:GRB_g_rate} is accurate to within a factor of few.
It should be noted that the numbers quoted above for the GRB rate density and $\gamma$-ray energy release are based on the assumption that GRB $\gamma$-ray emission is isotropic. If, as now commonly believed, the emission is confined to a solid angle $\Delta\Omega<4\pi$, then the GRB rate is increased by a factor $(\Delta\Omega/4\pi)^{-1}$ and the GRB energy is decreased by the same factor. However, their product, the energy generation rate, is independent of the solid angle of emission. 

\subsubsection{UHECR production rate}

The cosmic-ray spectrum flattens at $\sim10^{19}$~eV
\cite{fly,agasa}. There are indications that the spectral change
is correlated with a change in composition, from heavy to light
nuclei~\cite{fly,composition,HiResMIA} (see, however, \cite{WatsonComposition}). These characteristics suggest that the cosmic ray flux is
dominated at energies $< 10^{19}$~eV by a Galactic component of
heavy nuclei, and at UHE by an extra-Galactic source of protons.
Also, both the AGASA and Fly's Eye experiments report an
enhancement of the cosmic-ray flux near the Galactic disk at
energies $\le10^{18.5}$~eV,  but not at higher
energies~\cite{anisotropy,agasa1}. Fly's Eye stereo spectrum is
well fitted in the energy range $10^{17.6}$~eV to $10^{19.6}$~eV
by a sum of two power laws: A steeper component, with differential
number spectrum $J\propto \epsilon^{-3.50}$, dominating at lower energy,
and a shallower component, $J\propto \epsilon^{-2.61}$, dominating at
higher energy, $\epsilon>10^{19}$~eV. The data are consistent with the
steeper component being composed of heavy nuclei primaries, and
the lighter one being composed of proton primaries.

The observed UHECR flux and spectrum may be accounted for by a two component model \cite{BnW02}, with a Galactic component given by the Fly's Eye fit,$\frac{dn}{d\epsilon} ~\propto~ \epsilon^{-3.50}$,
and an extra-Galactic component characterized by a local ($z=0$) energy generation rate of extra-Galactic protons of
\begin{equation}
\epsilon_p^2\frac{d\dot{n}_p^{\rm CR}}{d\epsilon_p}=0.65\times 10^{44} {\rm
erg~Mpc^{-3}~yr^{-1}}. \label{eq:GRB_p_rate}
\end{equation}
Uncertainties in the absolute energy calibration of the experiments lead to uncertainty of $\approx20\%$ in this rate \cite{BnW02}.
The spectral index, 2, is that expected for acceleration
in sub-relativistic collisionless shocks in general, and in
particular for the GRB model discussed in
\S~\ref{sec:acceleration}. 
The model used in \cite{BnW02} is similar to that proposed in \cite{W95b}. The improved constraints on UHECR spectrum and flux provided by the recent observations of HiRes do not change the estimates given in \cite{W95b} for the energy generation rate and spectrum, Eq.~(\ref{eq:GRB_p_rate}), but reduce the uncertainties. 

Comparing eqs.~(\ref{eq:GRB_p_rate}) and (\ref{eq:GRB_g_rate}) we find that the observed UHECR flux and spectrum may be accounted for if GRBs produce high energy electrons and protons with similar spectra and rates. It should be emphasized here, that the ratio between the energy carried by relativistic electrons and protons in collisionless shocks is not known from basic principles. Moreover, as explained above, the estimate of the GRB gamma-ray energy generation rate is uncertain by a factor of a few. Thus, an exact match between the derived $\gamma$-ray and UHECR generation rates should not necessarily be expected. Given current uncertainties, the two rates should only be expected to be of the same order of magnitude. 

Finally, the following point should be mentioned. We have so far discussed the contribution of extra-Galactic GRBs to the UHECR flux. GRBs exploding within our Galaxy may also contribute to the Galactic cosmic-ray flux at lower energies, $>10^{15}$~eV \cite{MnU_Galactic}. However, in order for GRBs to make a significant contribution to the flux at $\sim10^{15}$~eV, they should produce more energy in $\sim10^{15}$~eV protons than they are inferred (based on their gamma-ray emission) to produce in high energy electrons \cite{Dermer_Galactic}.

\section{Predictions for high energy cosmic-ray experiments}
\label{sec:predictions}

\begin{figure}
\begin{center}
\includegraphics[width=3in]{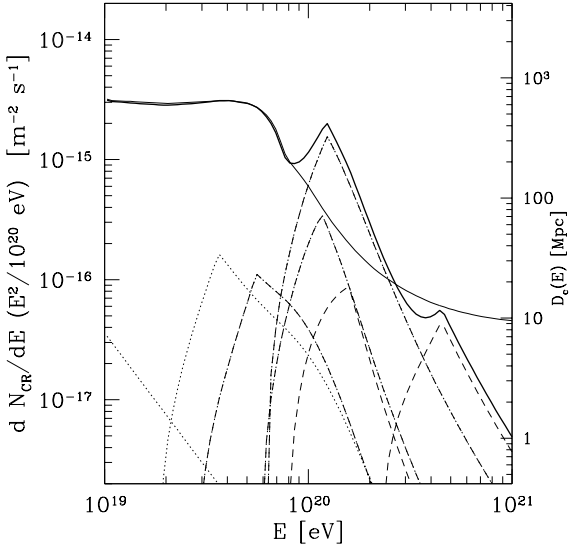}
\end{center}
\caption{Results of a Monte-Carlo realization of the bursting sources
model, with $\epsilon_c=1.4\times10^{20}$~eV: Thick solid line- overall 
spectrum in the realization;
Thin solid line- average spectrum, this
curve also gives $D_c(\epsilon_p)$;
Dotted lines- spectra of brightest sources at different energies.
}
\label{figNc}
\end{figure}

The initial proton energy, necessary to have an observed energy $\epsilon_p$,
increases with source distance due to propagation energy losses.
The rapid increase of the initial energy after it exceeds, due to
electron-positron production, the threshold for pion production effectively
introduces a cutoff distance, $D_c(\epsilon_p)$, 
beyond which sources do not contribute
to the flux above $\epsilon_p$. 
The function $D_c(\epsilon_p)$ is shown in Fig. \ref{figNc} (adapted from
\cite{MnW96}). Since $D_c(\epsilon_p)$ is a decreasing function of 
$\epsilon_p$, for
a given number density of sources 
there is a critical energy $\epsilon_c$, above which
only one source (on average) contributes to the flux. 
In the GRB model $\epsilon_c$ depends on the product of the burst 
rate $R_{GRB}$
and the time delay. The number of sources contributing, on average, 
to the flux at energy $\epsilon_p$ is \cite{MnW96}
\begin{equation}
N(\epsilon_p) = {4\pi\over 5} R_{GRB}D_c(\epsilon_p)^3 
\tau\left[\epsilon_p,D_c(\epsilon_p)\right]\quad,
\label{N}
\end{equation}
where $\tau(\epsilon_p,D)$, the spread in arrival time of protons of energy $\epsilon_p$ produced by a source at distance $D$ due to deflection by inter-galactic magnetic fields, is \cite{W95a,MnW96}
\begin{equation}
\tau(\epsilon_p,D)\approx
10^7\epsilon^{-2}_{p,20}D_{100}^2
\lambda_{\rm Mpc}B_{-8}^2\quad{\rm yr}.
\label{delay}
\end{equation}
Here $D=100D_{100}{\rm Mpc}$, $\lambda=1\lambda_{\rm Mpc}$~Mpc is the characteristic length scale over which the field direction varies, and $B=10^{-8}B_{-8}$~G is the characteristic field amplitude.
The average intensity resulting from all sources is
\begin{equation}
J(\epsilon_p) = \frac{1}{4\pi}R_{GRB} {d n_p\over d\epsilon_p} 
D_c(\epsilon_p)\quad,
\label{J}
\end{equation}
where $d n_p/d\epsilon_p$ 
is the number per unit energy of protons produced on average
by a single burst (this is the formal definition of $D_c(\epsilon_p)$).

The critical energy $\epsilon_c$, beyond which 
a single source contributes on average to
the flux,
\begin{equation}
{4\pi\over 5} R_{GRB}D_c(\epsilon_c)^3 \tau\left[\epsilon_c,
D_c(\epsilon_c)\right]=1\quad,
\label{Ec}
\end{equation}
depends on the unknown properties of 
the inter-galactic magnetic field, $\tau\propto B^2\lambda$. 
However, the rapid
decrease of $D_c(\epsilon_p)$ with energy near $10^{20}{\rm eV}$
implies that $\epsilon_c$ is only weakly dependent 
on the value of $B^2\lambda$. In The GRB model, the product $R_{GRB}\tau(D=100{\rm Mpc},\epsilon_p
=10^{20}{\rm eV})$
is approximately limited to the range $10^{-6}{\rm\ Mpc}^{-3}$ to
$10^{-2}{\rm\ Mpc}^{-3}$ (The lower limit is set by the requirement that 
at least a few GRB sources be present at $D<100$~Mpc, and the upper limit by 
the Faraday rotation bound 
$B\lambda^{1/2}\le10^{-8}{\rm G\ Mpc}^{1/2}$ \cite{Kronberg94}, and 
$R_{GRB}\le1/{\rm\ Gpc}^3{\rm yr}$). The corresponding range
of values of $\epsilon_c$ is 
$10^{20}{\rm eV}\le \epsilon_c<4\times10^{20}{\rm eV}$.

Fig. \ref{figNc} presents the flux obtained in one realization of
a Monte-Carlo simulation described by Miralda-Escud\'e \& Waxman 
\cite{MnW96} of the total
number of UHECRs received from GRBs at some fixed time. 
For each
realization the positions (distances from Earth) and
times at which cosmological GRBs occurred were randomly drawn, 
assuming an intrinsic proton
generation spectrum $dn_p/d\epsilon_p \propto \epsilon_p^{-2}$, and 
$\epsilon_c=1.4\times10^{20}{\rm eV}$. 
Most of the realizations gave an overall spectrum similar to that obtained
in the realization of Fig. \ref{figNc} when the brightest source of this 
realization (dominating at $10^{20}{\rm eV}$) is not included.
At $\epsilon_p < \epsilon_c$,
the number of sources contributing to the flux is very large, 
and the overall UHECR flux received at any
given time is near the average (the average flux is that obtained when 
the UHECR emissivity is spatially uniform and time independent).
At $\epsilon_p > \epsilon_c$, the flux will generally be much lower than the average,
because there will be no burst within a distance $D_c(\epsilon_p)$ having taken
place sufficiently recently. There is, however, a significant probability
to observe one source with a flux higher than the average.
A source similar to the brightest one in Fig. \ref{figNc}
appears $\sim5\%$ of the time. 

At any fixed time a given burst is observed in UHECRs only over a narrow
range of energy, because if
a burst is currently observed at some energy $\epsilon_p$ 
then UHECRs of much lower 
energy from this burst have not yet arrived,  while higher energy UHECRs
reached us mostly in the past. As mentioned above, for energies above the 
pion production threshold, 
$\epsilon_p\sim5\times10^{19}{\rm eV}$, the dispersion in arrival times of UHECRs
with fixed observed energy is comparable to the average delay at that
energy. This implies that
the spectral width $\Delta \epsilon_p$ of the source at a given time is of order
the average observed energy, $\Delta \epsilon_p\sim \epsilon_p$.
Thus, bursting UHECR sources should have narrowly peaked energy
spectra,
and the brightest sources should be different at different energies.
For steady state sources, on the other hand, the brightest
source at high energies should also be the brightest one at low
energies, its fractional contribution to the overall flux decreasing to
low energy only as $D_c(\epsilon_p)^{-1}$.
A detailed numerical analysis of the time dependent energy spectrum of 
bursting sources is given in \cite{Sigl97,Lemoine97}.

\section{Conclusions}
\label{sec:conclusions}

The main constraints that a relativistic wind (fireball) need to satisfy to allow proton acceleration to $>10^{20}$~eV are given by Eqs.~(\ref{eq:xiB}) and (\ref{eq:Gmin}): The magnetic field energy density $u_B$ should exceed a few percent of the relativistic electron energy density $u_e$, and the wind Lorentz factor $\Gamma$ should exceed $\approx10^2$. As explained in \S~\ref{sec:acceleration}, these constraints are independent of the details of the acceleration process, and are valid for any electromagnetic acceleration process. The similarity of these constraints and the constraints imposed on wind parameters, based on independent physical considerations, by $\gamma$-ray observations were the basis for the association of GRB and UHECR sources suggested in \cite{W95a}. We have pointed out in \S~\ref{sec:acc:Implications} that afterglow observations of GRBs imply a higher characteristic GRB luminosity than estimated based on $\gamma$-ray observations alone, thus relaxing the constraint of Eq.~(\ref{eq:xiB}) on $u_B/u_e$. Moreover, early optical and radio afterglow observations provide new constraints on wind parameters, implying large Lorentz factors, $\Gamma>10^2$ to $\Gamma>10^3$, and large magnetic field energy density in the fireball plasma, $u_B/u_e\sim1$. These constraints are consistent with those previously inferred from $\gamma$-ray observations, and with the constraints imposed by the requirement to allow proton acceleration to $>10^{20}$~eV. 

We have also pointed out, in \S~\ref{sec:rate}, that GRBs produce high energy electrons at a rate (and with a spectrum) similar to the rate (and spectrum) at which high energy protons should be produced in order to account for the UHECR spectrum and flux. Thus, the observed UHECR flux and spectrum may be accounted for if GRBs produce high energy electrons and protons at a similar rate and spectrum.

In \S~\ref{sec:predictions} we have pointed out that the local GRB rate implies that the rate of GRBs out to a distance from which most protons of energy exceeding $10^{20}$~eV originate, $\simeq90$~Mpc (see fig.~\ref{figNc}), is $\sim10^{-3}/$yr. The number of GRBs contributing to the observed flux at any given time, eq.~(\ref{N}) is given by the product of this rate and the spread in arrival time of protons, due to the combined effect of stochastic propagation energy loss and deflection by magnetic fields. This time spread, eq.~(\ref{delay}), may be as large as $10^7$~yr for $10^{20}$~eV originating at $90$~Mpc distance, implying that the number of GRBs contributing to the $>10^{20}$~eV flux at any given time may reach $\sim10^4$. The upper limit on the strength of the inter-galactic magnetic field, combined with the low local rate of GRB's, leads to unique predictions of the GRB model for UHECR production
\cite{MnW96,WnM96}, that may be tested with operating \cite{HiRes}, under-construction \cite{auger} and planned \cite{TA} large area UHECR detectors. In particular, a critical
energy is predicted to exist, $10^{20}{\rm eV}\le
E_c<4\times10^{20}{\rm eV}$, above which a few sources produce
most of the UHECR flux, and the observed spectra of these sources
is predicted to be narrow, $\Delta \epsilon/\epsilon\sim1$: The bright sources
at high energy should be absent in UHECRs of much lower energy,
since particles take longer to arrive the lower their energy.

\paragraph*{Acknowledgments.}
This research was partially supported by MINERVA and AEC grants.


\begin{thebibliography}{99}

\bibitem{Sigl}
  Bhattacharjee, P. \& Sigl, G. 2000, Phys. Rep. 327, 109.
\bibitem{NaganoWatson}
  Nagano, M. \& Watson, A. A. 2000, Rev. Mod. Phys. 72, 689.
\bibitem{pascos03} 
  Waxman, E. 2004, Pramana 62, 483 (Proc. PASCOS 03, Mumbai India; astro-ph/0310079).
\bibitem{W01rev} 
  Waxman, E. 2001, Lect. Notes Phys. {\bf 576}, 122-154 (astro-ph/0103186).
\bibitem{Fishman95}
  Fishman, G. J. \& Meegan, C. A. 1995, ARA\&A {\bf 33}, 415.
\bibitem{fireballs1}
  Piran, T. 2000, Phys. Rep. {\bf 333}, 529.
\bibitem{fireballs2}
  M\'esz\'aros, P. 2002, ARA\&A {\bf 40}, 137
\bibitem{fireballs3} 
  Waxman, E. 2003, Lect. Notes Phys. {\bf 598}, 393 (astro-ph/0303517).
\bibitem{W95a}
  Waxman, E. 1995, Phys. Rev. Lett. {\bf 75}, 386.
\bibitem{MnU95}
  Milgrom, M. \& Usov, V. 1995, Ap. J. {\bf 449}, L37.
\bibitem{Vietri95}
  Vietri, M. 1995, Ap. J. {\bf 453}, 883.
\bibitem{W95b}
  Waxman, E. 1995, Astrophys. J. Lett. {\bf  452}, L1.
\bibitem{W04}
  Waxman, E. 2004, Astrophys. J. {\bf 606}, 988.
\bibitem{PM04rev}
  M\'esz\'aros, P. et al. 2003, AIP Conference Proceedings, Vol. 727,
  125.
\bibitem{HalzenRev}
  Halzen, F. 2004, Nucl. Phys. {\bf B136}, 93.
\bibitem{non-relativistic}
  Blandford, R. \& Eichler. D. 1987, Phys. Rep. 154, 1.
\bibitem{relativistic}
  Waxman, E. 1997, Astrophys. J. 485, L5;
  Freedman, D. L. \& Waxman, E. 2001, Astrophys. J. {\bf 547}, 922; 
  Berger, I., Kulkarni, S. R. \& Frail, D. A. 2003 Astrophys. J. {\bf 590}, 379. 
\bibitem{rel-theory}
  Bednarz, J. \& Ostrowski, M. 1998, Phys. Rev. Lett. {\bf 80}, 3911;
  Kirk, J. K. et al. 2000, Phys. Rev. {\bf 542}, 235;
  Keshet, U. \& Waxman, E. 2004, to appear in Phys.Rev.Lett. (astro-ph/0408489).
\bibitem{arons}
  Arons, J. \& Tavani, M. 1994, Astrophys. J. Supp. 90, 797.
\bibitem{RnM98}
  Rachen, J. P., \& M\'esz\'aros,  P. 1998,  Phys. Rev. D {\bf 58},
  123005.
\bibitem{WnB-AG}
  Waxman, E., \& Bahcall, J. N. 2000, Ap. J. {\bf 541}, 707.
\bibitem{Dermer99}
  Dermer, C. D. 1999, A\&AS 138, 519.
\bibitem{Pelletier}
  Gialis, D. \& Pelletier, G. 2004. A\&A 425, 395; Gialis, D. \& Pelletier, G., submitted 
  to Ap. J. (astro-ph/0405547).
\bibitem{AG_ex_review}
  Kulkarni, S. R. {\it et al.} 2000,
  Proc. of the 5th Huntsville Gamma-Ray Burst Symposium
  (astro-ph/0002168).
\bibitem{Piran92}
  Piran, T. 1992, Ap. J. 389, L45.
\bibitem{MnP92}
  Mao, S. \& Paczy\'nski, B. 1992, Ap. J. 388, L45.
\bibitem{Bloom03}
  Bloom, J., Frail, D. A. \& Kulkarni, S.R., 2003, ApJ 594, 674.
\bibitem{ZKM03}
  Zhang, B., Kobayashi, S. \& M\'esz\'aros, P. 2003, ApJ 595, 950.
\bibitem{SR03}
  Soderberg, A. M. \& Ramirez-Ruiz, E. 2003, MNRAS 345, 854.
\bibitem{Draine00}
  Waxman, E.\ \& Draine, B.\ T. 2000, Ap. J. {\bf 537}, 796.
\bibitem{Galama98}
  Galama, T. J. 1998b, Nature 395, 670
\bibitem{Stanek03}
  Stanek, K. Z. et al. 2003, Ap. J. 591, L17
\bibitem{Hjorth03}
  Hjorth, J. et al. 2003, Nature 423, 847  
\bibitem{Bloom03a}
 Bloom, J. S. 2003, Proc. "Gamma Ray Bursts in the Afterglow Era, Third Workshop" (Rome, Sept 2002); 
 astro-ph/0303478.
\bibitem{GPW03}
  Guetta, D., Piran, T. \& Waxman, E. 2003, to appear in ApJ (astro-ph/0311488).
\bibitem{Band03}
  Band, D. L. 2003, Ap. J. 588, 945.
\bibitem{fly}
  Bird, D. J. {\it et al.} 1994, Astrophys. J. {\bf 424}, 491.
\bibitem{agasa}
  Hayashida N. {\it et al.} 1999, Astrophys. J. {\bf 522}, 225
  and astro-ph/0008102.
\bibitem{composition}
  Dawson, B. R., Meyhandan, R., and Simpson, K. M., 1998, Astropart. Phys. {\bf 9},
  331.
\bibitem{HiResMIA}
  Abu-Zayyad, T. et al. 2001, ApJ {\bf 557}, 686.
\bibitem{WatsonComposition} 
  Watson, A. A. Proc. XIII ISVHECRI: Pylos, Greece (astro-ph/0410514).
\bibitem{anisotropy}
  Bird, D. J. {\it et al.} 1998, Astrophys. J. {\bf 511}, 739.
\bibitem{agasa1}
  Hayashida N. {\it et al.} 1999a, Astropart. Phys. {\bf 10}, 303.
\bibitem{BnW02}
  Bahcall, J. N. \& Waxman, E. 2003, Phys. Lett. \textbf{B556}, 1
\bibitem{MnU_Galactic}
  Milgrom, M. \& Usov, V. 1996, Astropar. Phys. {\bf 4}, 365.
\bibitem{Dermer_Galactic}
  Wick, S. D., Dermer, C. D. \& Atoyan, A. 2004 Astropar. Phys. {\bf  21}, 125.
\bibitem{MnW96}
  Miralda-Escud\'e, J. \& Waxman, E. 1996, Ap. J. {\bf 462}, L59.
\bibitem{Kronberg94} 
  Kronberg, P. P. 1994, Rep. Prog. Phys.  {\bf 57}, 325.
\bibitem{Sigl97}
  Sigl, G., Lemoine, M. \& Olinto, A. V. 1997, Phys. Rev. {\bf D56}, 4470.
\bibitem{Lemoine97}
  Lemoine, M., Sigl, G., Olinto, A. V. \& Schramm, D. N. 1997, Ap. J. {\bf 486},
  L115.
\bibitem{WnM96}
  Waxman, E. \& Miralda-Escud\'e, J. 1996, Ap. J. {\bf 472}, L89.
\bibitem{HiRes}
  Abu-Zayyad, T., et al., 2004, Phys. Rev. Lett. {\bf 92}, 151101.
\bibitem{auger}
  Cronin, J.W. 1992, Nucl. Phys. B (Proc. Suppl.) 28, 313.
\bibitem{TA}
  Teshima, M. et al. 1992, Nucl. Phys. B (Proc. Suppl.) 28, 169.

\end{thebibliography}
\end{document}